\documentclass[aps,epsf,twocolumn,showpacs]{revtex4}
\usepackage{amsmath}
\usepackage{epsfig}

\begin{document}

\title{A comparative study of selected parallel tempering methods}

\author{A. Malakis}

\author{T. Papakonstantinou}

\affiliation{Department of Physics, Section of Solid State
Physics, University of Athens, Panepistimiopolis, GR 15784
Zografos, Athens, Greece}

\date{\today}

\begin{abstract}
We review several parallel tempering schemes and examine their
main ingredients for accuracy and efficiency. The present study
covers two selection methods of temperatures and several choices
for the exchange of replicas, including a recent novel all-pair
exchange method. We compare the resulting schemes and measure
specific heat errors and efficiency using the two-dimensional (2D)
Ising model. Our tests suggest that, an earlier proposal for using
numbers of local moves related to the canonical correlation times
is one of the key ingredients for increasing efficiency, and
protocols using cluster algorithms are found to be very effective.
Some of the protocols are also tested for efficiency and ground
state production in 3D spin glass models where we find that, a
simple nearest-neighbor approach using a local n-fold way
algorithm is the most effective. Finally, we present evidence that
the asymptotic limits of the ground state energy for the isotropic
case and that of an anisotropic case of the 3D spin-glass model
are very close and may even coincide.
\end{abstract}

 \maketitle

\section{Introduction}
\label{sec:1}

Monte Carlo (MC) sampling has increased dramatically our
understanding of the behavior of statistical mechanics
systems~\cite{Newman99,LandBind00}. Importance sampling, that is
the Metropolis \emph{et al.}~\cite{metro53} method and its
variants~\cite{Newman99,LandBind00} were, for many years, the main
tools in condensed matter physics, particularly for the study of
critical phenomena in Ising models. However, in many complex
systems, effective potentials have a complicated rugged landscape
with many minima and maxima which become more pronounced with
increasing system size. Thus, any reasonable MC sampling has to
overcome energy barriers and cross from one basin to another in
the state space in order to obtain a representative set of
configurations.

In order to overcome such problems, occurring mainly in the study
of first-order phase transitions, spin glasses and bio-molecules,
generalized ensembles have been developed with the two main
categories known as the entropic sampling method and the parallel
tempering sampling method~\cite{Newman99}. Parallel tempering
(PT), or replica exchange
ensembles~\cite{Swe86,Huku96,mariPL98,mari08,earl05,katz11} are
very effective alternatives for the study of spin
glasses~\cite{nishimori_book,BindKob05,EA_75,nishimori_1980,nishimori_1986,mcp_2d,doussal_harris_2,mcp_2,ozeki-ito,pala_sg,kawa_is,jorg,
hasen_mcp,campb_sg,hasen_sg,FG_1,bill-11,Huku98,ball00,Katz01,katz_pg},
protein folding~\cite{Sko01} and
biomolecules~\cite{Hans97,Sug99,Yam05}.

PT involves a mixture of MC moves at the individual temperatures
(local moves) with exchange attempts between different replicas
(swap moves). Local moves depend on the implemented local
algorithm, that is on the algorithm used at the individual
temperatures. Thus, a local move may be a spin-flip attempt
(Metropolis algorithm), or a spin flip to a new state (n-fold way
algorithm) or a cluster flip to a new state (Wolff algorithm). An
obvious question is how often we should perform the swap moves, or
how many local moves should intervene between swap moves. The
number of swap moves required to transfer any replica from the
highest (lowest) temperature to the lowest (highest), and vise
versa, defines the round-trip time and is one of the possible
global measures characterizing the efficiency of a PT protocol. It
is a measure indicating difficulties in the flow (bottleneck
effects) as a replica moves from the high to low temperature and
vise versa. Schemes minimizing the round-trip time are expected to
improve the general sampling efficiency of the PT protocols,
facilitating equilibration and transitions between multiple minima
in spin glasses.

There are many different ways to construct such a PT protocol and
the available freedom in choosing specific details makes their
comparison a challenging
job~\cite{mwa07,mach09,roma09,fiore11,Janus10}. Our motivation in
the present comparative study is to shed new light on aspects that
appear to be still undissolved. We attempt this by combining
several features and test the behavior of the resulting PT
protocols. We consider two methods for the selection of
temperature sequences and test their performance by varying the
numbers of local moves used at each temperature. One of these
methods is the simple constant acceptance exchange (CAE)
method~\cite{Sug99,earl05,katz11,bittner08}. Using this selection
method, Bittner \emph{et al.}~\cite{bittner08} have shown that if
the numbers of local moves are related to the canonical
correlation times, the resulting PT protocol appears to optimize
the round-trip time of replicas. The calculation of correlation
times needs additional costly preliminary runs, but the idea is of
theoretical interest, and in some cases an inferred moderate
approximation may be better than using an arbitrary
not-too-small/not-too-big empirical mixture of local and swap
moves.

The second method, for the selection of temperatures, was
introduced by Sabo \emph{et al.}~\cite{sabo08}. This method
requires a constant increase in entropy between successive
temperatures. This constant entropy increase (CEI) approach is
also supposed to optimize the performance of PT ensembles. The two
methods of selection produce temperature sequences that are more
concentrated in the temperature range where the specific heat has
a maximum, with the CEI method producing a more dense set close to
the maximum point. In systems with sharp specific heat peaks, the
difference between the corresponding protocols may be considerable
and one would like to know their relative efficiency and how this
may be influenced by other features of the protocols. We will
confirm that the proposal of Bittner \emph{et
al.}~\cite{bittner08} for using numbers of local moves related to
the canonical correlation times is the key ingredient for both
selection methods. Thus, using appropriate cluster algorithms for
the local moves one obtains very effective protocols optimizing
the round-trip time.

There is also great freedom in the choice of the exchange of
replicas, although most protocols use adjacent exchange attempts.
Even in the case of adjacent exchange moves, the performance of PT
is influenced by further details, such as the numbers of local
moves used between exchange attempts and the ordering or mixing of
the exchange attempts. Nonadjacent exchange attempts can be
incorporated in the PT schemes in a way that fully maintains
detailed balance. However, remote exchanges have in general very
small acceptance rates, and such schemes may well waste time in
unsuccessful attempts, without any statistically significant
increase in the sampling efficiency of the protocols. Such methods
of nonadjacent exchange moves have been also tried in constructing
PT protocols~\cite{neiro00} and will also be included in our tests
in this paper. An alternative procedure for overcoming the problem
of low acceptance rates for remote exchanges has been recently
adopted, by enforcing exchange of nonadjacent replicas using
kinetic MC methods~\cite{calvo05,brenner07}. The novel all-pair
exchange (APE) method of Brenner \emph{et al.}~\cite{brenner07} is
such a case. This method maintains detailed balance, at least in
determining the probabilities of generating exchange moves
(between adjacent or not adjacent replicas). We will consider this
method as a representative of kinetic MC methods and contrast its
accuracy and efficiency with the nearest neighbor ($nn$) PT
protocols. In particular, the combination of the APE method with a
Wolff (W) cluster algorithm~\cite{Newman99,Swendsen87} or a n-fold
way algorithm will be examined. This last algorithm is called also
in the literature BKL
algorithm~\cite{Newman99,bortz75,schulz01,malakis04b}, after
Bortz, Kalos, and Lebowitz~\cite{bortz75} who invented it. Other
names are continuous time MC or kinetic MC algorithm (see
discussion in Sec.~\ref{sec:2b}). The recent Infinite Swapping
Method~\cite{Platt11}, will not be considered in this paper. This
technique utilizes a symmetrization strategy, using all possible
($M!$) temperature (replica) permutations. The application and the
performance of this method to the problem of finding true ground
states of the 3D spin glass model is in our future interests. Our
numerical tests are carried out for the square Ising model with
linear size $L=50$, $N=L\times L$ lattice sites, and involve
accurate measurements for the specific heat errors and the
efficiency of the PT protocols. Furthermore, detailed tests are
carried out for efficiency and ground state production in 3D
spin-glass models, where we find efficient low-temperature choices
using, as a local algorithm, the n-fold way algorithm.

The rest of the paper is laid out as follows: In Sec.~\ref{sec:2a}
we give a brief description of two basic methods for selecting the
temperature sequences for PT sampling. The tested $nn$ and APE
exchange schemes are detailed in Sec.~\ref{sec:2b} and the rest of
the PT protocol details are defined in the elementary PT step in
Sec.~\ref{sec:2c}. In Sec.~\ref{sec:3} we define measures for the
specific heat errors and the efficiency and we present several
tests of the PT protocols on the 2D Ising model. Finally, in
Sec.~\ref{sec:4} we test the performance of some of the PT
protocols for ground state production on 3D spin-glass models. In
Sec.~\ref{sec:4a} we discuss the problem of ground states for the
3D Edwards-Anderson bimodal (EAB)
model~\cite{EA_75,nishimori_book}, while in Sec.~\ref{sec:4b} we
consider a variant of this, with spatially uniaxial anisotropic
exchange interactions and study the finite-size behavior of its
ground-state energy. Our conclusions are summarized in
Sec.~\ref{sec:5}.

\section{Parallel tempering schemes}
\label{sec:2}

\subsection{Selecting temperature sequences}
\label{sec:2a}

In constructing an accurate and efficient PT protocol, the optimum
selection of temperatures is still an open problem. There is a
rather large number of ideas that have been proposed in the last
decade to resolve this
question~\cite{katzg06,bittner08,kofk02,kon05,pred04,sabo08,neiro00,calvo05,brenner07}.
According to the approach followed by Katzgaber \emph{et
al.}~\cite{katzg06}, optimal temperatures correspond to a maximum
rate of round trips between low and high temperatures in
temperature space and can be obtained using a recursive
readjustment of temperatures. This feedback-optimized update
scheme, is a sophisticated and appealing method, but because of
its complexity, other simpler methods have been more often
implemented in comparative studies and applications of PT.

Among these simpler methods, the CAE method, when used with
appropriate number of sweeps between replica exchanges, has been
illustrated to produce a similar approach that optimize the
round-trip time~\cite{bittner08}. To obtain the temperatures
corresponding to a CAE rate $r$ we follow here
Ref.~\cite{bittner08}. Starting from a chosen lowest temperature,
adjacent temperatures are determined by calculating the acceptance
exchange rate from
\begin{equation}
\label{eq:1} R(1 \leftrightarrow 2) = \sum\limits_{E_1,E_2}
P_{T_1} (E_1) P_{T_2} (E_2) p(E_1,T_1 \leftrightarrow E_2,T_2),
\end{equation}
where $P_{T_i} (E_i)$ is the energy probability density function
for replica $i$ at temperature $T_i$ and
\begin{equation}
\label{eq:2} p(E_1,T_1 \leftrightarrow E_2,T_2) = min
[1,exp(\Delta \beta \Delta E)],
\end{equation}
is the PT probability to accept a proposed exchange of two
replicas, with $\Delta \beta=1/T_2-1/T_1$ and $\Delta E=E_2-E_1$.
Demanding that $R(1 \leftrightarrow 2)=r$ for all adjacent
replicas, we obtain the temperatures of the required CAE sequence
(from the above equations), provided that the energy probability
density functions (pdfs) are known, or can be reasonably well
approximated by some preliminary MC runs.

Although the exact density of states (DOS) and the energy pdf for
the square Ising model with linear size $L=50$ is
known~\cite{beale96}, we have used this information only for the
exact determination of specific heat errors and not for defining
the CAE temperature sequence. Instead, we apply a simple histogram
method~\cite{Swendsen87,Newman99} to find the energy pdfs at any
temperature, using a preliminary (Metropolis or PT) run in an
appropriate set of temperatures in the range of interest. Applying
then a recursive scheme we calculate the CAE sequence, and
repeating a Metropolis run at these temperatures we estimate their
canonical correlation times~\cite{Newman99}. This practice can be
applied to a general system for which the DOS is not known,
possibly using in the first preliminary run (especially in a spin
glass system) a PT protocol in an \emph{ad-hoc} reasonable set of
temperatures.

The second method of selection of the PT temperature sequence,
requires a constant increase in entropy between successive
temperatures~\cite{sabo08}. To describe this method we follow Sabo
\emph{et al.}~\cite{sabo08} and denote the $M$ temperatures of the
CEI sequence by $(T_m ; m=1,...,M)$ and the total increase in
entropy from $T_1$ to $T_M$ by $\Delta S$. Then the adjacent
temperatures are determined starting from the given $T_1$
successively from
\begin{equation}
\label{eq:3} \int_{T_m}^{T_{m+1}}dT \frac{C_u(T)}{T}=\frac{\Delta
S}{(M-1)},
\end{equation}
where the specific heat at any temperature can be calculated from
the above mentioned preliminary (Metropolis or PT) run and a
simple histogram method~\cite{Swendsen87,Newman99}. Repeating a
Metropolis run at these temperatures we also estimate the
corresponding canonical correlation times.

\begin{table*}
\caption{\label{tab:1} Temperature sequences for PT methods,
canonical correlation times in units of lattice sweeps, and
acceptance rates. The two cases shown are the constant acceptance
rate (columns 1,2 and 3) and the constant entropy increase
(columns 3,4 and 5) methods, as applied in a temperature range
close to the critical point of the $L=50$ square Ising model.}
\begin{ruledtabular}
\begin{tabular}{lcccccc}
   $$ &$CAE$ &$$ &$$ &$CEI$ &$$\\
      $T$ &$\tau$ &$r$ &$T$ &$\tau$ &$r$\\
\hline

                               1.9200  & 3.0 & 0.499  & 1.9200  & 3.0  & 0.361   \\
                               1.9669  & 3.0 & 0.500  & 1.9825  & 3.0  & 0.403   \\
                               2.0121  & 3.0 & 0.501  & 2.0375  & 3.6  & 0.428   \\
                               2.0557  & 3.9 & 0.499  & 2.0875  & 4.0  & 0.452   \\
                               2.0975  & 4.0 & 0.498  & 2.1325  & 4.8  & 0.480   \\
                               2.1377  & 5.0 & 0.501  & 2.1725  & 6.2  & 0.509   \\
                               2.1757  & 6.4 & 0.500  & 2.2075  & 10.0  & 0.551   \\
                               2.2115  & 9.5 & 0.501  & 2.2375  & 13.7  & 0.594   \\
                               2.2446  & 19.5 & 0.501  & 2.2625  & 25.6  & 0.607   \\
                               2.2751  & 32.6 & 0.501  & 2.2850  & 39.0  & 0.617   \\
                               2.3050  & 33.1 & 0.500  & 2.3075  & 31.3  & 0.601   \\
                               2.3374  & 19.1 & 0.500  & 2.3325  & 20.3  & 0.572   \\
                               2.3746  & 9.8 & 0.500  & 2.3625  & 11.2  & 0.531   \\
                               2.4167  & 6.3 & 0.500  & 2.4000  & 7.3  & 0.519   \\
                               2.4631  & 5.0 & 0.500  & 2.4425  & 5.6  & 0.489   \\
                               2.5134  & 4.1 & 0.500  & 2.4925  & 4.5  & 0.447   \\
                               2.5680  & 4.0 & 0.500  & 2.5525  & 4.0  & 0.434   \\
                               2.6268  & 3.8 & 0.500  & 2.6200  & 3.9  & 0.410   \\
                               2.6903  & 3.0 & .....  & 2.6975  & 3.0  & .....   \\

\end{tabular}
\end{ruledtabular}
\end{table*}

In Table~\ref{tab:1} we display, for the CAE and CEI selection
methods, the corresponding temperature sequences, their canonical
correlation times and the acceptance rates between adjacent
replicas. As usually, to fix the temperature scale we set (the
exchange interaction of the Ising model) $J/k_B=1$. The canonical
correlation times were estimated from the discrete form of the
energy autocorrelation function, following the method described in
Ref.~\cite{Newman99}, and are measured in units of lattice sweeps
($N$ Metropolis attempts). The temperature range used is
approximately centered around the pseudo-critical temperatures of
the specific heat and magnetic susceptibility of the $L=50$ square
Ising model, and includes the exact critical point. Both methods
produce temperature sequences that are more concentrated in the
temperature range where the specific heat has a maximum, with the
CEI method producing a more dense set close to the maximum point,
as can be seen from this table.

We point out here that, for the CAE selection, we start from a
given lower temperature $T_1=1.9200$ and proceed to find higher
temperatures, until a desired limit, with a given constant
acceptance rate. For the construction of Table~\ref{tab:1}, which
is used in our tests presented in Sec.~\ref{sec:3}, we have used
$r=0.5$ and the resulting higher temperature is $T_M=2.6903$ with
$M=19$. Our choice here for the CAE rate ($r=0.5$) follows
Ref.~\cite{bittner08}, and is somewhat arbitrary. One could also
use the value $r=0.3874$ recommended by~\cite{pred05}. However,
our test were repeated (on a lattice with linear size $L=20$)
using other values of the CAE rate, producing similar behavior. We
discuss again this point in Sec.~\ref{sec:3}, where we observe
that the CAE and CEI selections of temperatures show comparable
performance. Since we wish to compare the CAE and CEI schemes, we
apply the CEI procedure starting from the same $T_1=1.9200$, set
its final temperature $T_M=2.6975$, and use the same number of
replicas $M=19$. The small difference between the two higher
temperatures is due to the histogram data kept for the specific
heat in the preliminary run, since both sequences were obtained in
one unified algorithm.

Thus, the two schemes are defined approximately in the same
temperature range with the same number of replicas, a practice
that facilitates their comparison. The weaknesses or merits of the
two selection methods may also be related to the choice of lower
and higher temperature, the value of the constant acceptance rate
$r$, and thus will depend on the total number of replicas $M$. In
general, all the details of the protocols may influence the
round-trip time or the efficiency of the PT schemes. An
interesting recent example is that of the PT cluster algorithm
based on the CAE selection method presented by Bittner and
Janke~\cite{bittner11}. In their implementation, it was possible
to study the critical range of the 2D and 3D Ising model with a
rather small numbers of replicas. In the case 3D Ising model, it
was argued that a PT scheme using only $M=4$ replicas, for lattice
sizes $L=4-80$, was adequate to cover the critical range. Finally,
these authors used in their study the more generous approach with
$M=7-21$ replicas, for lattice sizes $L=4-80$.

\subsection{Nearest neighbor and all-pair exchange schemes}
\label{sec:2b}

Mixing local MC attempts at individual temperatures (local
attempts) with exchange attempts between different replicas (swap
attempts) is the essential procedure in PT. It is this feature
that enables an ergodic walk in temperature space, transferring
information between the highest and lowest temperatures. We
provide in this section short descriptions of alternatives for the
ordering of swap attempts, first for the $nn$ exchange protocols
and then for PT protocols using all-pair exchange(APE) methods.
The rest of the details of an elementary PT step are described in
the next section.

In a $nn$ exchange protocol only adjacent exchange attempts are
proposed. There are $M-1$ different $nn$ proposals which may be
uniquely denoted by the lowest temperature index $i=1,...,M-1$.
Following Brenner \emph{et al.}~\cite{brenner07}, we denote the
replica configuration before a swap attempt by $A=\left
\{x_1,x_2,...,x_i,x_{i+1},...,x_M\right \}$, where $x_i$ is the
replica at the temperature $T_i$. Then, a $nn$ exchange is denoted
by $A \rightarrow B$, where $B=\left
\{x_1,x_2,...,x_{i+1},x_i,...,x_M\right \}$. The
acceptance/rejection rule of an exchange attempt corresponds to an
acceptance rate $P_{acc}(A \rightarrow B)$, which is usually given
by the Metropolis form of the swap attempt $p(x_i \leftrightarrow
x_{i+1})$, as specified in Eq.~(\ref{eq:2}). Since the generation
of the various $nn$ exchange attempts (proposals) proceeds with
equal probabilities, $P_{gen}(A \rightarrow B)=P_{gen}(B
\rightarrow A)=1/(M-1)$, the swap attempts satisfy the detailed
balance condition $P(A)P_{gen}(A \rightarrow B)P_{acc}(A
\rightarrow B)=P(B)P_{gen}(B \rightarrow A)P_{acc}(B \rightarrow
A)$. The product probability distribution
$P(A)=\rho(x_1)\rho(x_2)...\rho(x_M)$ is stationary with respect
to the swap attempts. We now specify four choices
[$(nn)_a$,$(nn)_b$,$(nn)_c$, and $(nn)_d$] for the ordering of the
$nn$ proposals that will be tested in Sec.~\ref{sec:3}. In
$(nn)_a$, a random permutation (say: $j_1,...,j_{M-1}$) is
generated, from the set $i=1,...,M-1$, and this permutation is
used in an exchange swap cycle of $M-1$ proposals. Thus, swap
moves are organized in cycles of $M-1$ proposals, and as explained
in the next section, such a swap cycle may be used in defining the
unit of time of an elementary PT step. In $(nn)_b$, the lowest
temperature index ($i$) is randomly chosen from the set
$i=1,...,M-1$ and thus multiple exchanges of the same pair are
allowed in the swap cycle. In $(nn)_c$, the sequence of proposals
is fully deterministic, consisting of two swap sub-cycles,
starting from the odd proposals $i=1,3,...$, in increasing order,
and continuing with the even proposals $i=2,4,...$. Finally, in
$(nn)_d$, the odd and even sub-sequences are randomly permuted
before starting the odd and following with the even swap
sub-cycles.

Next, we consider PT protocols using APE methods, in which
adjacent or not adjacent replicas may be exchanged $(x_i
\leftrightarrow x_j)$. In these methods the number of possible
proposals ($A \rightarrow B$) is $M(M-1)/2$, $A=\left
\{x_1,x_2,...,x_i,...,x_j,...,x_M\right \}$ and  $B=\left
\{x_1,x_2,...,x_j,...,x_i,...,x_M\right \}$. In the simplest case,
for each proposal a pair $(x_i,x_j)$ is randomly chosen from the
set of all $M(M-1)/2$ different pairs. Thus, the generation of
exchange attempts proceeds with equal probabilities, $P_{gen}(A
\rightarrow B)=P_{gen}(B \rightarrow A)=1/[M(M-1)/2]$, and if the
acceptance/rejection rule follows the Metropolis form, with an
acceptance rate $p(x_i \leftrightarrow x_j)$, then the swap
attempts satisfy the detailed balance condition. This simple APE
method will be included in our tests and is denoted, in the
sequel, as $APE_M$, whereas the method of Brenner \emph{et
al.}~\cite{brenner07} will be denoted by $APE_B$. As pointed out
in the introduction, methods attempting remote exchange moves have
been also tried in constructing PT protocols~\cite{neiro00} and
will be included in our tests in Sec.~\ref{sec:3}. Remote
exchanges may be thought as replacing several adjacent swaps to a
single move. However, the $APE_M$ method suffers very small
acceptance rates, and no significant increase in the sampling
efficiency of the protocols should be expected.

Finally, we discuss the so-called kinetic MC methods and present
details of the APE methods proposed by Calvo~\cite{calvo05} and
Brenner \emph{et al.}~\cite{brenner07}. In statistical physics,
particularly in simulation studies of Ising models, the kinetic MC
method is better known as the n-fold way algorithm or BKL
algorithm~\cite{Newman99,bortz75,schulz01,malakis04b}. In
describing this algorithm, we shall follow the original
Ref.~\cite{bortz75} and borrow from the terminology in Section
$2.4$ of Ref.~\cite{Newman99}. In the traditional MC simulation
(for instance Metropolis algorithm) we use an acceptance/rejection
rule in every MC attempt, and the system may stay in the same
state for some time $\Delta t$. In a kinetic MC algorithm, we
enforce the system to select a new state, but we also introduce a
time-step which corresponds to a varying length (stochastic
variable), depending on how long we expect the system to remain in
its current state before moving to a new one in the traditional MC
method~\cite{Newman99}. The averaging process for any observable
becomes a time average and the values of the observable are
weighted by the corresponding time-steps (divided, at the end of
the measuring process, by the total time), while the total time
variable $t$ is incremented by $\Delta t$. The selection of a new
state assumes an appropriate set of probabilities and proceeds as
follows: Let the current state of the system be $\mu$ and denote
by $r_j\equiv r(\mu\rightarrow\nu)$ the acceptance rates
(acceptance probabilities) for all possible $K$ transitions
($j=1,2,...,K$) to new states from the current state. Draw a
random number $0<R \leq 1$ and select the new state $\nu$,
corresponding to transition $i$, if $Z_{i-1} < RZ_K \leq Z_i$,
where $Z_m=\sum\limits_{j=1}^m\ r_j$. With the help of the
cumulative function $Z_m$, the new state $\nu$ is selected with
probability $P_i=r_i/Z_K$, proportional to the acceptance
probability $r(\mu\rightarrow\nu)$. The $P_j$ are the selection
probabilities of the kinetic MC algorithm. The time intervals
$\Delta t$, have to be recalculated at each step. They can be
obtained as average life-times~\cite{schulz01}, from the values of
$r(\mu\rightarrow\nu)$, and $\Delta t\propto
{Z_K}^{-1}$~\cite{Newman99,bortz75,schulz01,malakis04b}. Note also
that, in an equivalent way, the selection of a new state $i$ can
proceed with the condition stated in terms of a cumulative
function $Q_m$ obtained from the selection probabilities. In this
case, the condition reads as $Q_{i-1} < R \leq Q_i$, where
$Q_m=\sum\limits_{j=1}^m\ P_j$.

These are the main ingredients of the a kinetic MC algorithm. In
some cases, the set of all transitions can be classified to a
small number of $n$ classes (n-fold way algorithm) and the method
becomes very efficient, but in general the recalculation of all
transition probabilities in each step, is the obvious drawback of
the method. In Ref.~\cite{bortz75}, the n-fold way algorithm is
described in detail for the square lattice Ising model [periodic
boundary conditions (PBC)] in a nonzero field, using the $n=10$
different classes, corresponding to the energy changes under a
spin flip. For the zero field square lattice (with $N$ sites and
PBC) Ising model, $n=5$ classes cover the spin flip energy changes
$\Delta E_j=8-4(j-1)$, where the index numbering the classes is
$j=5-z$ and $z=4,3,2,1,0$ is the number of nearest neighbor spins
having the same sign with the spin to be
flipped~\cite{bortz75,schulz01,malakis04b}. In this case, the
statistical weight for the selection of a class is the sum of the
acceptance rates of all spins in the class and takes the form:
$r^{cl}_j=N_jA_j$, where $N_j$ are the current populations of
spins ($\sum\limits_j\ N_j=N$) and $A_j=min [1,exp(-\beta \Delta
E_j)]$ is the corresponding (Metropolis) acceptance rate of any
spin in the class. The class selection proceeds as outlined above,
using the cumulative function $Z_m=\sum\limits_{j=1}^m\ r^{cl}_j$
and the average life-times are given by $\Delta t
=N{Z_n}^{-1}$~\cite{schulz01}. The spin to be flipped is then
chosen randomly from the selected class and the spin flip is
enforced. This is a simple and efficient n-fold way algorithm, a
kinetic MC reorganization of the original Metropolis
algorithm~\cite{bortz75}, with transitions that follow detailed
balance. The method has been used successfully to simulate
thermodynamic equilibrium of various types of Ising like models,
using, as mentioned earlier, an appropriate time averaging
process. This version of n-fold way algorithm, is implemented, as
a local algorithm, in Sec.~\ref{sec:3} for the square lattice
Ising model, and in Sec.~\ref{sec:4} for the 3D (cubic) spin glass
models ($n=7$).

The APE method proposed by Calvo~\cite{calvo05}, is an attempt to
adjust the above ideas of the kinetic MC to the PT swapping
procedure. Following the terminology of Brenner \emph{et
al.}~\cite{brenner07}, let $\Phi$ denote the set of all
macrostates (replica configurations) $B=\left
\{x_1,x_2,...,x_j,...,x_i,...,x_{M-1},x_M\right \}$ reachable from
the current macrostate $A=\left
\{x_1,x_2,...,x_i,...,x_j,...,x_{M-1},x_M\right \}$ by an adjacent
or not adjacent pair exchange $(x_i \leftrightarrow x_j)$. This
set describes all possible transitions $A \rightarrow B$ to a new
state in a kinetic MC scheme. Yet, Calvo also includes the
possibility not to perform any exchange (event $j=0$) to which the
acceptance probability $P_{acc}(A \rightarrow A)=1$ is
attributed~\cite{calvo05}. Thus, the PT swapping procedure of
Calvo~\cite{calvo05}, involves, besides the $K=M(M-1)/2$
transitions to new states of the form $A \rightarrow B\neq A$
($j=1,2,...,K$), also the event $A \rightarrow A$ ($j=0$). This is
an unconventional use of the kinetic MC method and the selection
probabilities of the events are given by~\cite{calvo05,brenner07}
\begin{equation}
\label{eq:4} P_j=P_{acc}(A \rightarrow C)/[1+\sum\limits_{M \in
\Phi} P_{acc}(A \rightarrow M)],
\end{equation}
where $C$ is $A$ or $B$. The event $i$ is selected, and is
enforced, from the set of $j=0,1,2,...,K$, using the cumulative
function $Q_m$ obtained from the selection probabilities. The
condition is $Q_{i-1} < R \leq Q_i$, where
$Q_m=\sum\limits_{j=1}^m\ P_j$. As pointed out by
Calvo~\cite{calvo05}, the attribution of a residence time (average
life-time) is inconvenient in PT schemes and has been replaced in
his method by including the rejection $j=0$ event. However, this
is not a self-consistent use of the kinetic MC method for
estimating thermodynamic equilibrium. Therefore,
Calvo~\cite{calvo05} suggests, that the exchange moves (enforced
without the use of time weights) could be considered as extra
moves that are not directly involved in the averaging process. In
a subsequent paper, Brenner \emph{et al.}~\cite{brenner07} pointed
out a further inconsistency of the method. The set of macrostates
$\Phi$ reachable from $A$ is not, in general, the same with the
set of macrostates $\Psi$, reachable from $B$. Thus, the
generation probability [inverse of the denominator in
Eq.~(\ref{eq:4})] $P_{gen}(A \rightarrow B)=[1+\sum\limits_{M \in
\Phi} P_{acc}(A \rightarrow M)]^{-1}$ for the transition $A
\rightarrow B$, will in general be different from the generation
probability $P_{gen}(B \rightarrow A)=[1+\sum\limits_{L \in \Psi}
P_{acc}(B \rightarrow L)]^{-1}$ for the transition $B \rightarrow
A$. As a consequence the detailed balance condition $P(A)P_{gen}(A
\rightarrow B)P_{acc}(A \rightarrow B)=P(B)P_{gen}(B \rightarrow
A)P_{acc}(B \rightarrow A)$ is not met.

In order to overcome this inconsistency, Brenner \emph{et
al.}~\cite{brenner07} proposed a revision of the above method. The
generation probabilities are now replaced by~\cite{brenner07}
\begin{align}
&P_{gen}(A \rightarrow B)=\nonumber\\
&1/{max \left \{\sum\limits_{M \in \Phi} P_{acc}(A \rightarrow
M),\sum\limits_{L \in \Psi} P_{acc}(B \rightarrow L)\right \}}
\label{eq:5}
\end{align}
and the event generation probabilities satisfy now the desired
relation $P_{gen}(A \rightarrow B)=P_{gen}(B \rightarrow
A)$~\cite{brenner07}. This condition makes the scheme consistent
with detailed balance, in contrast with the method originally
proposed by Calvo~\cite{calvo05}. The PT swapping procedure of
Brenner \emph{et al.}~\cite{brenner07}, involves explicitly only
the $K=M(M-1)/2$ transitions of the form $A \rightarrow B\neq A$
($i=1,2,...,K$) and the selection probabilities for the kinetic MC
method are given by
\begin{equation}
\label{eq:6} P_j=P_{acc}(A \rightarrow B)P_{gen}(A \rightarrow B),
\end{equation}
where $P_{acc}(A \rightarrow B)$ is the corresponding PT
acceptance rate $p(x_i \leftrightarrow x_j)$, as specified in
Eq.~(\ref{eq:2}), and $P_{gen}(A \rightarrow B)$ is given by
Eq.~(\ref{eq:5}). The selection of the state (macrostate) from the
set of $j=1,2,...,K$ states proceeds again, using the cumulative
function $Q_m$ obtained from the selection probabilities. The
condition is $Q_{i-1} < R \leq Q_i$, where
$Q_m=\sum\limits_{j=1}^m\ P_j$. The algorithm of Brenner \emph{et
al.}~\cite{brenner07} does not include explicitly, in the set of
selection probabilities, the possibility not to perform any
exchange. However, this may happen, in the direction of the less
favorable Boltzmann exchange~\cite{brenner07}, since $P_{gen}$ is
a maximum and thus in some cases $Q_K=\sum\limits_i P_i <1$. In
such cases, no new state is selected from the condition $Q_{i-1} <
R \leq Q_i$, where $Q_m=\sum\limits_{j=1}^m\ P_j$ if $R>Q_K$. But
again, the method proposed by Brenner \emph{et
al.}~\cite{brenner07} does not use any time weights (residence
times or average life-times) for the MC averaging process, as is
usually done in a kinetic MC algorithm when it is used for the
estimation of thermodynamic equilibrium properties. This is a
common problem of the above methods, and it may be crucial for the
estimation of any observable in thermodynamic equilibrium. The
accuracy of the resulting PT protocols has not been tested before
and it can not be guessed beforehand.

We will examine this last APE method in combination with several
local algorithms, including the Metropolis, the n-fold way, and
the cluster Wolff algorithm. As mentioned earlier, we shall use
the notation $APE_B$ to refer to the described all-pair exchange
method of Brenner \emph{et al.}~\cite{brenner07} and the notation
$APE_M$ for the simple APE method using for the swap probabilities
the acceptance/rejection rule of Metropolis form given in
Eq.~(\ref{eq:2}). The APE method proposed by Calvo~\cite{calvo05},
will not be included in the presentation of our tests, and we only
mention here that, as we have verified, this method is slightly
inferior to the APE method of Brenner \emph{et
al.}~\cite{brenner07}. A further obvious drawback of Brenner
\emph{et al.}~\cite{brenner07} method is the costly recalculation
of all generation probabilities in each exchange step for the
application of Eq.~(\ref{eq:5}) and Eq.~(\ref{eq:6}). Note that,
before selecting a particular swap event in the kinetic MC method
of Brenner \emph{et al.}~\cite{brenner07}, all proposals $A
\rightarrow B$ are considered and for each of them both sums in
the denominator of Eq.~(\ref{eq:5}) are calculated from the known
acceptance rates. For this reason, the $APE_B$ PT protocol needs
considerable more CPU time. This additional time can be reduced by
observing that, the $APE_B$ protocol does not create with
significant probability distant exchanges. More details are
provided in the discussions of Sec.~\ref{sec:3}.

\subsection{The elementary parallel tempering step and the local algorithms}
\label{sec:2c}

We now define a PT step (PTS), as the elementary MC step used for
the recording (measuring or averaging) process during an
independent MC run. A PTS may consist of one or several swap
cycles of $M-1$ replica exchange proposals and we use this
definition also in the case of the APE method, so that our unit is
the same for all protocols to be tested. After each exchange
attempt of the swap cycle, all replicas attempt a number of local
moves (spin flips or cluster moves) at their respective
temperatures. The number of these local moves is in general chosen
to depend on the temperature, and is denoted by $n(T_i)$. The
total number of local moves at any temperature of the PT protocol
($T$), in a swap cycle, is $N_{local}(T)=(M-1)n(T)$. The swap
cycle is thus, the above described mixture of standard MC and
replica exchange attempts. Without loss of generality, we define
the PTS to be just one such swap cycle, and we differentiate
between various protocols by varying the number of locals moves.
It is convenient to set also $N_{local}(T)=f(T)\tau(T)N$, where
$N=L^D$ is the number of lattice sites, $\tau(T)$ are the
canonical correlation times and $f(T)$ are factors that facilitate
the adaption of the numbers of local moves $n(T)$ in a style
depending on the application and/or the behavior of the system.
$N_{local}(T)=N$, corresponds to the usual lattice sweep.

The above defines our MC unit for the recording process. We note
that, no recording is attempted during a first disregarding or
equilibration part of the simulation, and an adequate number of
PTSs, $t_{eq}$, is used for this part. Then, we use a large
number, $t_{av}$, of PTSs for the recording or averaging part of
the simulation. For the local standard MC moves, the Metropolis
algorithm, the n-fold way algorithm and the cluster Wolff
algorithm will be implemented. Furthermore, a large number $N_r$
of independent MC runs is used in our tests. This repetition makes
more reliable the measures of accuracy and efficiency illustrated
in our tests. It is also desirable to use approximately the same
CPU time for the different PT schemes that are compared. In this
way, we hope to obtain an objective assessment of their behavior.
Since the local algorithms obey different time complexities, we
will adapt the parameters $N_{local}(T)$, $t_{eq}$, $t_{av}$, and
$N_r$ in a way that makes the algorithms (almost) equivalent in
CPU time.

\section{Error and efficiency measures: Tests on the 2D Ising model}
\label{sec:3}

In this section we present tests, carried out by using the L=50
square ferromagnetic Ising system with PBC. A variety of PT
protocols will be implemented, obtained by combining features
mentioned in the previous sections. Using the exact
DOS~\cite{beale96}, we calculate the values of the specific heat
at the PT protocol temperatures and define the following
error-measures

\begin{equation}
\label{eq:7}
\epsilon(T_i)=[C_{exact}(T_i)-C_{PT}(T_i)]/C_{exact}(T_i),
\end{equation}

\begin{equation}
\label{eq:8} \bar{\epsilon}=\sum\limits_{i=1}^M\epsilon(T_i)/M,
\end{equation}

\begin{equation}
\label{eq:9} \hat{\epsilon}=\sum\limits_{i=1}^M|\epsilon(T_i)|/M,
\end{equation}

\begin{equation}
\label{eq:10} \epsilon^*=\max(|\epsilon(T_i)|),
\end{equation}

\begin{figure}[htbp]
\includegraphics*[width=9 cm]{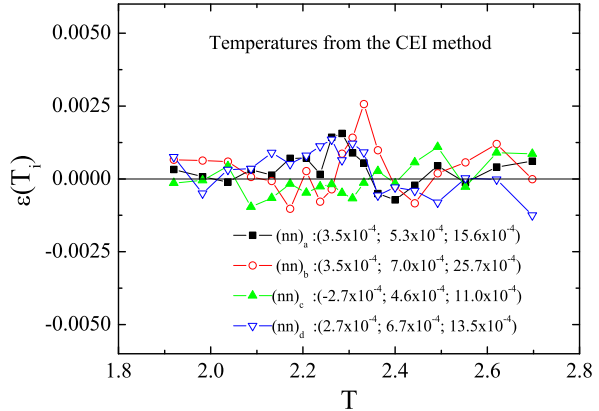}
\caption{\label{fig:1} (Color online) Specific heat errors for the
nearest neighbor ($nn$) PT protocols. The CEI selection of
temperatures, as given in Table~\ref{tab:1}, has been used. The
error measures, given in parenthesis in all panels, are
respectively $\bar{\epsilon}$, $\hat{\epsilon}$ and $\epsilon^*$.}
\end{figure}

Our first test concerns the $nn$ PT protocols and in particular
the four cases $(nn)_a$, $(nn)_b$, $(nn)_c$, and $(nn)_d$,
described in Sec.~\ref{sec:2b}. In Fig.~\ref{fig:1}, we illustrate
their errors $\epsilon(T_i)$, and on the panel (in parenthesis) we
give the corresponding error-measures, namely $\bar{\epsilon}$,
$\hat{\epsilon}$, and $\epsilon^*$, as defined above. In each of
the presented cases, we have used $N_r=200$ independent MC runs
with $t_{eq}=3N$, $t_{av}=15N$ and $N_{local}(T)=N$ and the
Metropolis algorithm for the local moves. It is evident from the
excellent accuracy obtained in all cases, that there is not any
noticeable and statistically significant difference between the
four choices, rather it appears that they all obey a good mixing
of the exchange attempts in a long run. Therefore, hereafter we
will only use the $(nn)_a$ PT protocol and vary the other
ingredients of the schemes.

\begin{figure}[htbp]
\includegraphics*[width=9 cm]{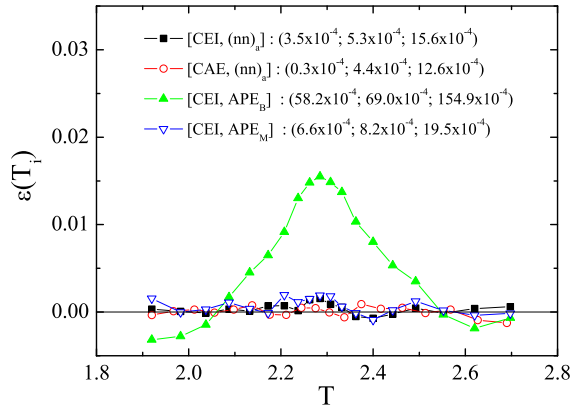}
\caption{\label{fig:2}(Color online) Specific heat errors for four
different PT protocols. The first two use the nearest-neighbor
protocol $(nn)_a$, while the last two use the all-pair exchange
protocols of Brenner \emph{et al.}~\cite{brenner07} and Metropolis
(subsection~\ref{sec:2b}). The method for the selection of
temperatures and the resulting error measures are indicated in the
notation on the panel.}
\end{figure}

We now attempt to observe whether the two selections of
temperatures produce any significant difference. Also, we test the
$nn$ against the APE protocols. In Fig.~\ref{fig:2}, we illustrate
the error behavior of two $nn$ and two APE protocols using both
the CEI and CAE selections as indicated on the panel. The details
of the illustrated PT schemes [$N_r, t_{eq}, t_{av},
N_{local}(T)$], are the same as the ones specified above. The $nn$
protocols perform excellently when compared with the APE
protocols. The CAE and CEI selections give comparable accuracy.
The $APE_B$ protocol suffers very large errors in the specific
heat that are more pronounced close to the specific heat maximum.
Apparently, this erratic behavior is a reflection of the problem
mentioned earlier. The $APE_B$ method does not use any time
weights for the MC averaging process, and therefore is, in
general, unreliable for the estimation of thermodynamic
equilibrium properties. On the other hand, the $APE_M$ protocol,
which is a standard PT protocol attempting also distant
(Metropolis) exchanges, shows a good error behavior. We should
also point out here, that the $APE_B$ runs need almost double CPU
time, due to the calculations needed for the application of
Eqs.~(\ref{eq:5}) and (\ref{eq:6}). However, this additional time
can be greatly reduced by observing that, the $APE_B$ protocol
does not create with significant probability distant exchanges,
and thus can be restricted only to significant exchanges.

In order to observe the diffusion behavior of the PT protocols, we
define for each temperature $T_i$, the fraction of replicas which
have visited one of the two extremal temperatures most recently.
In our implementations we assign labels $up$ or $down$ to each
replica if its most recent visit to one of the two extremal
temperatures is $T_M$ or $T_1$, respectively. The label of an $up$
replica remains unchanged if the replica returns to $T_M$, but
changes to $down$ upon its first visit to $T_1$. Similarly, the
label of a $down$ replica remains unchanged if the replica returns
to $T_1$, but changes to $up$ upon its first visit to $T_M$. For
each temperature $T_i$, we record histograms $n_{up}(T_i)$ and
$n_{down}(T_i)$, which are incremented by one after a swap attempt
involving an $up$ or $down$ replica at $T_i$ respectively.

\begin{figure}[htbp]
\includegraphics*[width=9 cm]{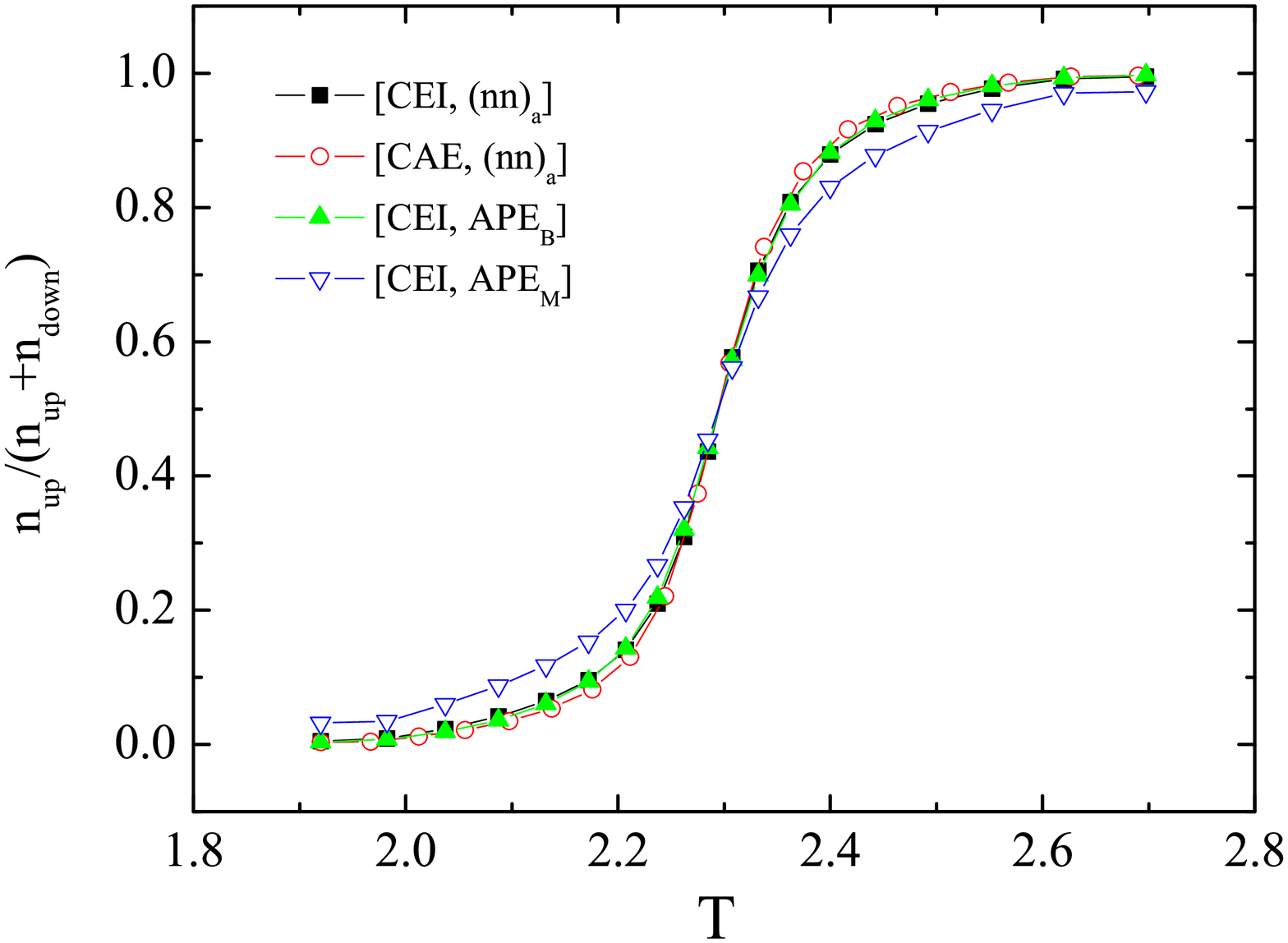}
\caption{\label{fig:3}(Color online) Diffusion behavior of the PT
protocols. The diffusion fraction
$n_{up}(T_i)/(n_{up}(T_i)+n_{down}(T_i))$, as defined in the text,
for the four PT protocols of Fig.~\ref{fig:2}.}
\end{figure}

The diffusion fraction $n_{up}(T_i)/(n_{up}(T_i)+n_{down}(T_i))$
is illustrated in Fig.~\ref{fig:3}, for the PT protocols of
Fig.~\ref{fig:2}. Apparently, the same definition has been used in
Ref.~\cite{bittner08} and a similar one (with $up$ replaced by
$down$ and vise versa) in Ref.~\cite{katzg06}. All four cases in
Fig.~\ref{fig:3} show very similar diffusion behavior with a sharp
decline close to the critical point (specific heat maximum). Thus,
a significant increase of efficiency of kinetic MC
methods~\cite{calvo05,brenner07}, that attempt to enforce remote
exchanges, is not verified from the illustrated diffusion
behavior. Figure~\ref{fig:3} also illustrates the similarity in
the diffusion behavior between the two selection methods. The
larger concentration of replicas of the CEI method, close to the
maximum specific heat point, does not produce any significant
difference in the diffusion behavior of the protocol. We observe
again comparable performance for the CAE and CEI selection rules.
This observation, may be related to the finding of
Ref.~\cite{pred05}(see Figure 1., and the related discussion in
section V. Conclusions of~\cite{pred05}). Namely, moderate
variations in the acceptance rate ($r$) do not produce significant
changes in the efficiency of PT protocols, provided the acceptance
rates remain in the range of optimum performance.

\begin{figure}[htbp]
\includegraphics*[width=9 cm]{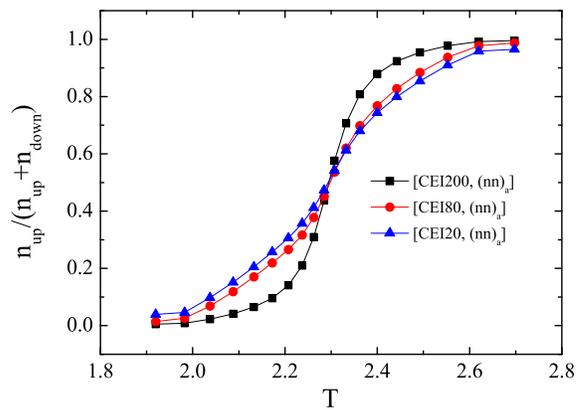}
\caption{\label{fig:4}(Color online) This figure illustrates the
changes induced in the behavior of the diffusion fraction by
varying the numbers of local moves in accordance with the
canonical correlation times, as proposed by Bittner \emph{et
al.}~\cite{bittner08}. The three protocols use the CEI selection
and correspond to $N_{local}(T)=N$, $N_{local}(T)=0.25\tau(T)N$
and $N_{local}(T)=\tau(T)N$. The corresponding numbers of
independent MC runs are $N_r=200$, $80$, and $N_r=20$ as indicated
in the notation on the panel.}
\end{figure}

\begin{figure}[htbp]
\includegraphics*[width=9 cm]{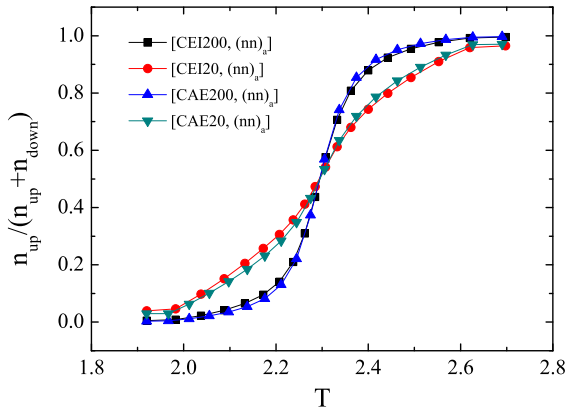}
\caption{\label{fig:5}(Color online) The first two CEI protocols
are the same with those illustrated in that figure, while the
other two are the corresponding CAE protocols.}
\end{figure}

Our next test, illustrated in Fig.~\ref{fig:4}, verify the
important observation by Bittner \emph{et al.}~\cite{bittner08}.
By varying the numbers of local moves, in accordance with the
canonical correlation times, we observe strong changes induced in
the behavior of the diffusion fraction. The three protocols
illustrated use the CEI selection with $t_{eq}=3N$ and
$t_{av}=15N$, and correspond to $N_{local}(T)=N$,
$N_{local}(T)=0.25\tau(T)N$ and $N_{local}(T)=\tau(T)N$. The
numbers of independent MC runs are $N_r=200$, $80$, and $N_r=20$
as indicated in the notation on the panel of this figure
($[CEI200,(nn)_a]$,$[CEI80,(nn)_a]$, and $[CEI20,(nn)_a]$,
respectively), and thus in total the schemes require approximately
the same CPU time (from Table~\ref{tab:1}, the mean of the
correlation times is approximately $10$).
Furthermore, in Fig.~\ref{fig:5} we reproduce the
$[CEI200,(nn)_a]$ and $[CEI20,(nn)_a]$ cases of the previous
figure with the corresponding CAE protocols. This comparison
supplements the crucial observation that the key ingredient,
improving the efficiency of the protocols, is the proper choice of
numbers of local moves.

An alternative way to measure the efficiency of the PT protocols,
is to observe global aspects of the statistics of the numbers of
exchange attempts required to transfer any replica from the
highest (lowest) temperature to the lowest
(highest)~\cite{brenner07}. Let, $u_j$ ($d_j$) denote the average
numbers of exchange attempts required for the corresponding
transfer, averaged over the PTSs of a long independent run $j$
($j=1,2,...,N_r$) and over the $M$ different replicas of the
protocol. These quantities are strongly fluctuating in the
ensemble of $N_r$ independent runs and is more convenient to
illustrate the behavior of the running averages of their
combination
\begin{equation}
\label{eq:11} [(u+d)/2]_k=\sum\limits_{j=1}^k((u_j+d_j)/2)/k,
\end{equation}
where $k=1,2,...,N_r$, and we will also refer to the ratio of
their running averages $[u]_k/[d]_k$.

\begin{figure}[htbp]
\includegraphics*[width=9 cm]{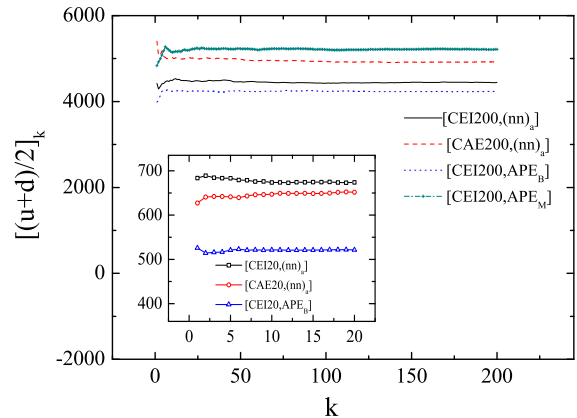}
\caption{\label{fig:6}(Color online) Illustration of the
efficiency of the PT protocols using the measures defined in
Eq.~(\ref{eq:11}). The four protocols of the main panel use
$N_{local}(T)=N$, while the three PT protocols in the inset use
$N_{local}(T)=\tau(T)N$.}
\end{figure}

Figure~\ref{fig:6} provides a clear illustration of the efficiency
of the PT protocols. We observe the strong influence of the
numbers of local moves on the efficiency measure defined in
Eq.~(\ref{eq:11}). This verifies again, that the choice of the
numbers of local moves, related to the canonical correlation
times, is the decisive ingredient of all PT protocols for
increasing efficiency. Thus, our tests support the proposal made
earlier by Bittner~\emph{et al.}~\cite{bittner08} and also show
that the influence of the selection method is of minor importance.
Furthermore, the influence of the $APE_B$ method is also marginal
on the efficiency and we should keep in mind that this method
suffers from larger specific-heat errors, as illustrated in
Fig.~\ref{fig:2}. The ratio of the corresponding running averages
$[u]_k/[d]_k$ is not equal to one, but depends on several details
of the protocols and in particular on the selection of the highest
and lowest temperature. Although, it appears that the efficiency
is higher when this ratio is close to one, the differences between
the protocols using the same numbers of local moves are rather
small.

\begin{figure}[htbp]
\includegraphics*[width=9 cm]{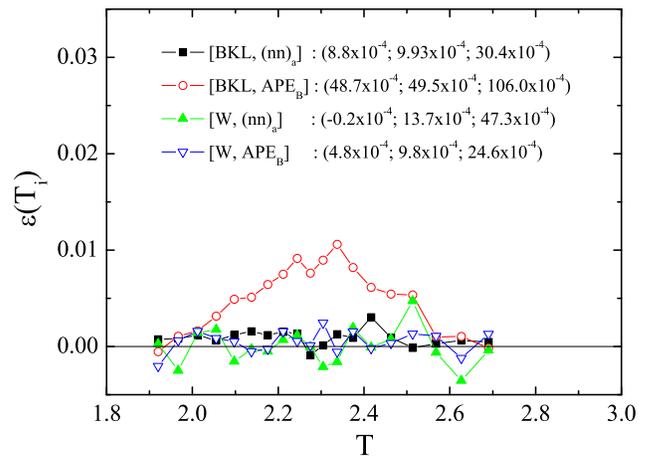}
\caption{\label{fig:7}(Color online) Specific heat errors for the
PT protocols shown in the panel and discussed in more details in
the text. The CAE selection of temperatures, as given in
Table~\ref{tab:1}, has been used in all four cases.}
\end{figure}

\begin{figure}[htbp]
\includegraphics*[width=9 cm]{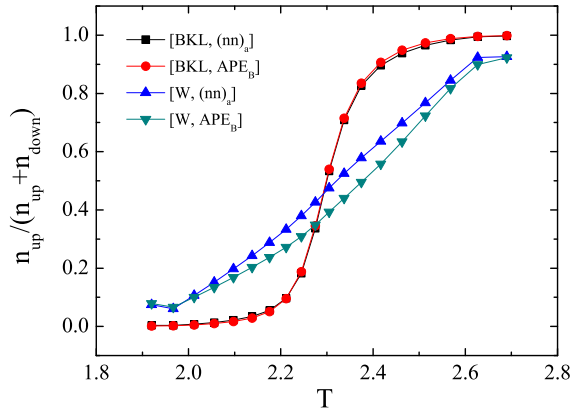}
\caption{\label{fig:8}(Color online) Illustration of the diffusion
fraction for the PT protocols shown in the panel. Again, the CAE
selection of temperatures, as given in Table~\ref{tab:1}, has been
used in all four cases.}
\end{figure}

\begin{figure}[htbp]
\includegraphics*[width=9 cm]{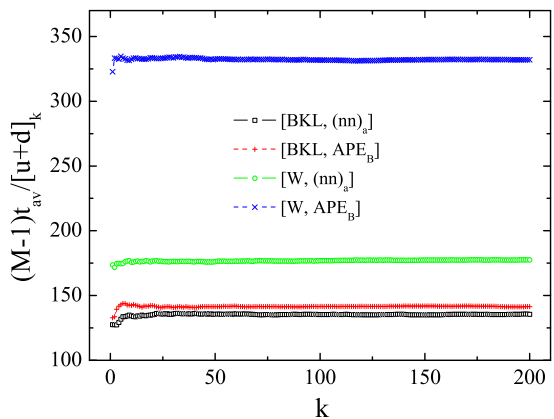}
\caption{\label{fig:9}(Color online) Illustration of running
averages of the quantity $(M-1)t_{av}/[u+d]_k$, which is the mean
number of round-trips of a replica during one independent run. The
larger running averages of these numbers correspond to the most
efficient protocol which is now definitely the $[W,APE_B]$ PT
scheme.}
\end{figure}

The above findings raise questions regarding the importance of
introducing cluster algorithms for the local moves in the PT
schemes, since as is well known, cluster algorithms have very
small dynamical exponents. Therefore, it should be expected that
the implementation of such algorithms, for the local moves, will
increase the efficiency of the protocols. Naturally, we now
consider schemes that use for the local moves two further
alternatives, besides the Metropolis algorithm. These two
alternatives are: (i) the Wolff cluster algorithm (denoted in the
figures as W), which is known for its small dynamical exponent,
and (ii) the one-spin flip algorithm known as the n-fold way or
BKL algorithm (denoted in the figures as BKL) described in
Sec.~\ref{sec:2b}.

In order to simplify our presentation, we consider now only the
CAE selection method and we omit the initials CAE from our
notation. Thus, the protocols of interest are now denoted by
$[BKL,(nn)_a]$, $[BKL,APE_B]$, $[W,(nn)_a]$, and $[W,APE_B]$,
corresponding to $nn$ exchange and the APE method of Brenner
\emph{et al.}~\cite{brenner07}. Their behavior is illustrated in
the following three figures.

In particular, Fig.~\ref{fig:7} illustrates the specific heat
error behavior of these four PT protocols. Comparing the two
$APE_B$ protocols of Fig.~\ref{fig:7} with the $APE_B$ protocols
of Fig.~\ref{fig:2} which are using a local Metropolis algorithm,
we observe clear improvements in the behavior of the illustrated
error measures. The improvement is substantial in the case of the
$[W,APE_B]$ protocol and moderate in the case of the $[BKL,APE_B]$
protocol. In order to appreciate these improvements we specify the
rest of the details of the protocols. The number of independent MC
runs is again $N_r=200$ in all four cases. For the $[BKL,(nn)_a]$
and $[BKL,APE_B]$ protocols we use again $t_{eq}=3N$,
$t_{av}=15N$, and $N_{local}(T)=(M-1)n(T)=0.216N$, which
corresponds to $n(T)=30$ BKL spin flips before each swap move.
This choice makes the time requirement of the BKL protocols
approximately equivalent to the corresponding Metropolis
protocols. As mentioned earlier, the $APE_B$ PT protocols requires
more time, due to the extra calculations needed for the
application of Eqs.~(\ref{eq:5}) and~(\ref{eq:6}), before each
swap move. Yet, we can achieve approximately the same time
requirements for this protocol, by restricting the possible remote
exchanges up to forth order. We note that the sum of probabilities
of orders $5,6,7$, and $8$ is only $0.074\%$ and the higher orders
never occur. The probabilities of the first four orders of
exchanges are $71.01\%$, $23.17\%$, $5.03\%$, and $0.72\%$ for
first, second, third and fourth neighbor exchanges respectively.
The error behavior and the efficiency of the protocol is very
weakly influenced, by this restriction, and thus it is more
reasonable to use this restriction than to use small parameters
for the $APE_B$ protocols. In the cases using the Wolff algorithm
for the local moves, we have used $t_{eq}=3N/50$, $t_{av}=N$, and
$N_{local}(T)=(M-1)n(T)=0.0216N$, which corresponds now to
$n(T)=3$ Wolff cluster flips before each exchange attempt. The
above adjustments, for the Wolff PT protocols, are essentially
reflecting the small dynamical exponent of the Wolff algorithm, in
an attempt to construct a protocol having approximately the same
time requirements with the Metropolis PT protocols.

Figure~\ref{fig:8} illustrates the behavior of the diffusion
fraction of the four protocols. One can now observe that while the
BKL cases behave very similarly to the Metropolis case shown in
Fig.~\ref{fig:3}, the Wolff cases show a behavior resembling the
Metropolis protocols using numbers of local moves analogous to the
canonical correlation times of Figs.~\ref{fig:4} and~\ref{fig:5}.
In fact the changes induced in the behavior of the diffusion
fraction are now more spectacular. Furthermore, Fig.~\ref{fig:9}
provides a demonstration that the combination of the $APE_B$
exchange method combined with a local Wolff algorithm is now the
best choice for increasing the efficiency of the PT scheme. Since,
the used numbers of PTSs, $t_{av}$, are different for the BKL and
Wolff protocols we have plotted in Fig.~\ref{fig:9} the running
averages of the quantity $(M-1)t_{av}/[u+d]_k$. This defines the
mean number of round trips of a replica during one independent run
and is larger for the most efficient protocol which is now clearly
the $[W,APE_B]$ PT scheme.

Thus, taking also into account the observation that, the
implementation of the Wolff algorithm removes the problem with the
large specific heat errors, we may declare here that, under some
conditions, the $APE_B$ PT method may provide a respectable and
efficient PT protocols. This appears as a justification of the
suggestion of Calvo~\cite{calvo05}: the exchange moves could be
considered as extra moves that are not directly involved in the
averaging process. An explanation for the generally large
specific-heat errors of the $APE$ schemes, may be sought in
problems coming from the omission of appropriate residence times
and possible strong disturbances caused by remote exchanges
enforced by the $APE$ exchange methods. Apparently, these seem to
be cured by the very fast restoration of equilibrium that takes
place after the Wolff local moves. On the other hand, the moderate
improvement of the $[BKL,APE_B]$ protocol, compared with the
corresponding $[M,APE_B]$ in Fig.~\ref{fig:2}, may be due to the
use, in local level, of appropriate average life-times, as is
usually done in a BKL
implementation~\cite{Newman99,bortz75,schulz01,malakis04b}. The
above observations provide clues for improving the all-pair
exchange methods, such as those proposed by Calvo~\cite{calvo05}
and Brenner \emph{et al.}~\cite{brenner07}. Our tests on the Ising
model illustrate some of the merits and weaknesses of these
schemes. Of course, the real power of all PT methods, should be
checked in rare-event problems in which the performance of
conventional MC methods can become unreliable.

\section{Ground states of 3D spin-glass models}
\label{sec:4}

We proceed to test the efficiency of PT protocols for the
production of ground states (GS) in 3D spin glass models. It is
well known that finding the GS of spin glass system in $D=3$ is an
NP-complete problem~\cite{houd01,hart04} and there exist a large
number of heuristic algorithms developed in order to tackle this
outstanding problem~\cite{hart04,boe01,pal96,hart97,katzGS}. In a
recent paper, Roma~\emph{et al.}~\cite{roma09} have concluded that
PT is comparable to the performance of the more powerful
heuristics. In particular, they have concentrated on the
estimation of the minimum number of PTSs needed to achieve a true
GS with a given probability. In their study they considered the
Edwards-Anderson model in 2D and 3D with bimodal and Gaussian bond
disorder. We will consider the 3D Edwards-Anderson bimodal (EAB)
model~\cite{EA_75,nishimori_book} and an anisotropic 3D EAB model.

We will first discuss, in Sec.~\ref{sec:4a}, the standard
(isotropic) 3D EAB model and observe the relative performance of
some of the presented PT protocols. Then, in Sec.~\ref{sec:4b}, we
shall consider a variant of this model, with spatially uniaxial
anisotropic exchange interactions. This model has been studied
recently~\cite{anis_2}, and its phase diagram has been estimated.
For this case we will provide results for its finite size behavior
of the GS energy, comparing our estimates with the isotropic case.
The anisotropic spin glass model~\cite{anis_1,anis_2}, is defined
by the Hamiltonian
\begin{equation}
\label{eq:12} H=-\sum_u\sum_{\langle ij
\rangle_u}J_{ij}^us_{i}s_{j},
\end{equation}
where the exchange interactions are uncorrelated quenched random
variables, taking the values $\pm J^{xy}$ on the $xy$ planes and
the values $\pm J^{z}$ on the z axis. The bimodal distribution of
$J_{ij}^u$ takes the general form
\begin{equation}
\label{eq:13} P(J_{ij}^u) =
p_u\delta(J_{ij}^u+J^u)+(1-p_u)\delta(J_{ij}^u-J^u),
\end{equation}
where $u$ denotes the $z$ axis $(u=z)$ or the $xy$ planes
$(u=xy)$, $J^u$ denotes the corresponding exchange interaction
strength and $p_u$ are the probabilities of two neighboring spins
($ij$) having antiferromagnetic interaction. The standard
isotropic EAB model, corresponds to $J^z=J^{xy}=J(=1)$ and
$p_z=p_{xy}$. In the following we consider the production of GS
for the isotropic $(p_z=p_{xy}=0.5)$ and anisotropic
$(p_z=0,p_{xy}=0.5)$ cases.

\subsection{Ground states of the 3D EAB model: Further tests of PT protocols}
\label{sec:4a}

In this section, we consider the standard isotropic 3D EAB model
on a cubic lattice of linear size $L=6$. For this model,
Roma~\emph{et al.}~\cite{roma09} have addressed the question of
whether it is more efficient, in order to find a true GS, to use
large running times or several independent runs ($t_{av}$ and
$N_r$ in our notation) for each realization of the disorder,
called henceforth sample. We note that, due to some differences in
defining the PTS, our notation is not identical with that in
Ref.~\cite{roma09}, but the analogies are obvious and we will also
use $N_s$ to denote the number of different samples. The collapse
example in Fig. 2b. of Ref.~\cite{roma09} shows clearly that
increasing the number of PTSs ($t_{av}$), has approximately the
same effect with an analogous increase in the number of
independent runs ($N_r$) for each sample. Accordingly the
production of a true GS depends on the product $N_rt_{av}$. Thus,
in what follows we demonstrate the performance of some of the
previous PT protocols, using $N_r=1000$ and $N_r=100$ for the
number of independent runs, and adjusting $t_{av}$ and the other
parameters of the protocols in a way that enabled us to compare
schemes requiring approximately the same CPU time. We will
implement single spin-flip local algorithms, since an efficient
cluster algorithm for the production of GSs in the spin glass
problem is not available. Primarily, we wish here to observe the
relative efficiency and dependence of the PT schemes on the local
algorithms, on the exchange method used, and also on the selection
method of temperatures for the PT process. These issues are
distinguishing our comparative approach from that of Roma~\emph{et
al.}~\cite{roma09}, who have concentrated mainly on estimating the
running times necessary for generating true GS with a given
probability and not on differentiating among various PT recipes,
using different local and swap moves. However, their estimated
times have guided our tests, and also our study of the GS energy
of the mentioned anisotropic variant of the EAB model, considered
in the next subsection.

\begin{table*}
\caption{\label{tab:2} Temperature sequences obtained by averaging
over $N_s=50$ samples the individual T-sequences. The individual
T-sequences are obtained by relatively short runs, as described in
the text. Sequences for two values of the acceptance rate $r$ of
the CAE method, together with the corresponding CEI method
sequences are shown.}

\begin{ruledtabular}
\begin{tabular}{lcccccccc}
    \multicolumn{2}{c}{$CAE(r=0.1):M=5$} & \multicolumn{2}{c}{$CEI:M=5$}& \multicolumn{2}{c}{$CAE(r=0.5):M=11$} &  \multicolumn{2}{c}{$CEI:M=11$}\\
\cline{1-2}  \cline{3-4}  \cline{5-6} \cline{7-8}
      $T$ &$r$ &$T$ &$r$ &$T$ &$r$ &$T$ &$r$\\
\hline

                               0.5000  & 0.097  & 0.5000  & 0.033  &0.5000  & 0.481  & 0.5000  & 0.211 \\
                               0.8868  & 0.103  & 1.0200  & 0.130  &0.6893  & 0.486  & 0.7800  & 0.382 \\
                               1.2269  & 0.112  & 1.3550  & 0.222  &0.8934  & 0.498  & 0.9600  & 0.511 \\
                               1.5940  & 0.105  & 1.6750  & 0.144  &0.9746  & 0.498  & 1.1150  & 0.489 \\
                               2.0235  & .....  & 2.0350  & .....  &1.1137  & 0.501  & 1.2550  & 0.537 \\
                               &&&                                 &1.2553  & 0.516  & 1.3900  & 0.548 \\
                               &&&                                 &1.4018  & 0.501  & 1.5250  & 0.519 \\
                               &&&                                 &1.5560  & 0.507  & 1.6600  & 0.637 \\
                               &&&                                 &1.7191  & 0.510  & 1.8000  & 0.622 \\
                               &&&                                 &1.8973  & 0.498  & 1.9500  & 0.552 \\
                               &&&                                 &2.0950  & .....  & 2.1100  & ..... \\
\end{tabular}
\end{ruledtabular}
\end{table*}

In Table~\ref{tab:2}, we give the temperature sequences
(T-sequences) obtained by averaging the individual T-sequences
over $50$ samples of the 3D EAB model for a lattice size $L=6$.
The individual T-sequences were obtained using relatively short
runs, by the histogram method outlined in Sec.~\ref{sec:2a}.
Sequences for two values of the acceptance rate $r$ of the CAE
method, together with the corresponding CEI method sequences, are
shown. The T-sequences, displayed in Table~\ref{tab:2}, correspond
to the values $r=0.1$ and $r=0.5$, involving $M=5$ and $M=11$
temperatures respectively. Since, each sample has its own
T-sequences, the illustrated averaged T-sequences will be, in any
case, only a rough approximation for each sample. Thus, short PT
runs, using an \emph{ad-hoc} reasonable set of temperatures, were
found to be adequate for each of the $50$ samples used to
construct the table. Of course, one can improve this approximation
by increasing the PT running times ($t_{av}$) for recording the
histograms (of energy and specific heat), but this will produce
non-significant changes in the final averaged T-sequences.
However, some defects of this short time approximation can be
observed in the fluctuation of the exchange rate in the CEI method
in the second part of the table for the case $r=0.5$.

We have used a set of $N_s=1000$ samples, in order to test the
performance of several PT protocols. The local algorithms used are
the Metropolis (M) and the BKL (n-fold way) algorithms. The swap
moves are carried out by the $(nn)_a$ and the $APE_B$ methods. In
Table~\ref{tab:3} we summarize these tests in three groups. The
CAE T-sequence of $r=0.1$ with $M=5$ temperatures has been used in
the first and second group of four schemes, while both the CAE and
CEI T-sequences, corresponding to the case $r=0.5$ and $M=11$
temperatures has been used in the third group of five schemes. In
an obvious notation, in each case we give the abbreviation for the
T-sequence, then the acronym of the local algorithm and finally
the method of PT exchange. Thus, $[CAE,M,(nn)_a]$ denotes the PT
protocol based on the CAE sequence, using Metropolis algorithm for
the local moves and the $(nn)_a$ method for the swap moves, while
$[CEI,BKL,APE_B]$ denotes the PT protocol based on the CEI
sequence, using the BKL algorithm for the local moves and the
$APE_B$ method for the swap moves.

\begin{table*}
\caption{\label{tab:3} Sample averages ($[P]$) and minimum
($P_{min}$) probabilities of generating a true GS for various PT
protocols. In each case, the notation indicates the implementation
of the Metropolis (M) or the BKL algorithms for the local moves,
and the implementation of the $(nn)_a$ or the $APE_B$ methods for
the PT exchanges. Finally, we have indicated the employed
T-sequences (CAE or CEI selection method) and give in the table
the main parameters of the PT protocols.}
\begin{ruledtabular}
\begin{tabular}{lccccccc}
    $CAE(r=0.1):M=5$  &$N_r$ &$n=N_{local}/(M-1)$ &$t_{eq}$ &$t_{av}$ &$[P]$ &$P_{min}$ \\
\hline

                               $[CAE,M,(nn)_a]$  & 1000 &$54$ &$108$ &$216$ &0.8205 &0.026\\
                               $[CAE,M,APE_B]$  & 1000 &$54$ &$108$ &$216$ &0.8302 &0.024\\
                               $[CAE,BKL,(nn)_a]$  & 1000 &$16$ &$108$ &$216$ &0.9324 &0.035\\
                               $[CAE,BKL,APE_B]$  & 1000 &$16$ &$108$ &$216$ &0.9398 &0.039\\
\hline
    $CAE(r=0.1):M=5$  &$N_r$ &$n=N_{local}/(M-1)$ &$t_{eq}$ &$t_{av}$ &$[P]$ &$P_{min}$ \\
\hline

                               $[CAE,M,(nn)_a]$  & 100 &$54$ &$108$ &$2160$ &0.9769 &0.05\\
                               $[CAE,M,APE_B]$  & 100 &$54$ &$108$ &$2160$ &0.9787 &0.04\\
                               $[CAE,BKL,(nn)_a]$  & 100 &$18$ &$108$ &$2160$ &0.9921 &0.09\\
                               $[CAE,BKL,APE_B]$  & 100 &$18$ &$108$ &$2160$ &0.9929 &0.11\\
\hline
    $CAE(r=0.5):M=11$  &$N_r$ &$n=N_{local}/(M-1)$ &$t_{eq}$ &$t_{av}$ &$[P]$ &$P_{min}$ \\
\hline

                               $[CAE,M,(nn)_a]$  & 100 &$21$ &$64$ &$1271$ &0.9856 &0.12\\
                               $[CEI,M,(nn)_a]$  & 100 &$21$ &$64$ &$1271$ &0.9818 &0.10\\
                               $[CAE,BKL,(nn)_a]$  & 100 &$5$ &$64$ &$1271$ &0.9936 &0.15\\
                               $[CEI,BKL,(nn)_a]$  & 100 &$5$ &$64$ &$1271$ &0.9917 &0.14\\
                               $[CEI,BKL,APE_B]$  & 100 &$5$ &$64$ &$1271$ &0.9930 &0.15\\

\end{tabular}
\end{ruledtabular}
\end{table*}

For a particular sample, a large number of independent PT runs
($N_r$) is carried out and some of these runs ($n_j$) successfully
find a true GS. Thus, $P_j=n_j/N_r$ is the probability of reaching
a GS for the $jth$ sample. This probability will depend on the
details of the PT protocols, but is also strongly dependent on the
particular sample. It is well known, that easy and hard samples
exist with very different behavior (see for instance Fig. 15. of
~\cite{roma09}), and one expects that in a large ensemble of
samples the hardest samples have the smallest $P_j$. The minimum
of this probability $P_{min}$, in a given ensemble, will therefore
give us an indication of the performance of the PT protocol for
the hard samples, while the sample average
$[P]=\sum\limits_{j=1}^{N_s}P_j/N_s$, reflects the average global
performance of the PT protocol. The introduction of these
probabilities for the production of true GSs, enables us now to
discuss the entries of Table~\ref{tab:3} and compare the PT
schemes.

For each sample $j$ of the set $N_s=1000$ samples, used for the
construction of Table~\ref{tab:3}, we denote the GS energy per
site $u_{6,j}$, indicating also in the notation the lattice size
($L=6$). The sample average ($u_6$) of the GS energy per site, for
the particular set of samples, was found to be
$u_6=\sum\limits_{j=1}^{N_s}u_{6,j}/N_s=-1.77026$. This appears as
an exact result, for the chosen set of samples, since true GSs
have been found for all samples (with probability almost $1$). To
verify that true GSs were found for each sample, we have used also
longer runs, than those in Table~\ref{tab:3}, by a factor of $4$.
Note here, that according to Ref.~\cite{roma09}, the probability
of true GS production for running time $t=2\times10^5$ is of order
$0.999$, and that the recording times ($N_rt_{av}$), used for
Table~\ref{tab:3}, are of the same order. Thus, using the longer
runs and observing no difference in the above estimate $u_6$, we
conclude that true GSs have been found for all samples.

As mentioned earlier, in each of the three groups of the examples,
shown in Table~\ref{tab:3}, care has been taken to adjust the PT
parameters in a way that corresponds to the same CPU time within
each group and reflects the different time requirements of the
protocols involving local BKL or Metropolis moves. The CPU time of
the second and third group are approximately the same while the
CPU time of the first group is larger by a factor $1.88$, mainly
because in the first group, the disregarding equilibration part
($t_{eq}$) is comparable to the recording part ($t_{av}$) of the
protocols. We note here that, in all our implementations, the
search for grounds states and the rest of the recording processes
are carried out only in the recording part ($t_{av}$) of the runs.
The fine adjustments of parameters, needed in order to achieve
approximately the same CPU time in each group, were fixed by short
preliminary runs.

From Table~\ref{tab:3}, we observe the superiority of the BKL
algorithm compared to the Metropolis algorithm for the local
moves. This is reflected, within each group, in the values of the
global sample averaged probability $[P]$, and it is more
pronounced in the first group, where the averaged probability is
not very close to $1$. The superiority of the BKL algorithm, is
also reflected, within each group, in the values of the minimum
probabilities $P_{min}$, that concern the behavior of hardest
samples. The observed here superiority is not a surprise from an
algorithmic point of view, since n-fold way updates have been the
basis for previous attempts in searches for GSs of spin
glasses~\cite{hart04,boe01,paolo01}. The all-pair exchange $APE_B$
method improves rather marginally the production process of GSs.
Finally, comparing the CAE and CEI cases in the third group, we
notice that the results for the CAE T-sequence are slightly better
than the corresponding results for the CEI T-sequence.

In the first and second group of Table~\ref{tab:3} only the
results corresponding to the CAE T-sequence are shown. These CAE
results are again better than the corresponding results for the
CEI T-sequence (not included in the table). The requirement of an
exchange rate $r=0.1$ yields sequences with only $M=5$
temperatures. In this case, the lowest temperature exchange rate
for the CEI method is too low (see Table~\ref{tab:2}). This could
be the source of the superiority of the CAE selection method.
However, as illustrated in Table~\ref{tab:3}, even in the case of
an exchange rate $r=0.5$, the CAE results are slightly better.
Finally, comparing the probabilities of the first group with the
other two groups, one can see the strong dependence of the sample
averaged probability $[P]$ on the running times $t_{av}$, as
should be expected and pointed out by Roma~\emph{et
al.}~\cite{roma09}.

The main conclusion of this section is the observation that, among
several PT recipes tested, the PT protocol using for the local
moves the BKL algorithm and a T-sequence obtained by the CAE
method is superior to the other tested protocols. This conclusion
was verified by more tests not presented here for brevity.
Furthermore, we have tried to observe the effect of using larger
numbers of local moves on the  probability measures $P_{min}$ and
$[P]$. In particular the run $[CAE,BKL,(nn)_a]$, of the second
group in Table~\ref{tab:3}, was repeated using the choices
($n=4\times18; t_{av}=2160$), ($n=2\times18; t_{av}=2\times2160$),
and ($n=18; t_{av}=4\times2160$). Note that, these cases are
almost equivalent in CPU time. The resulting probabilities were
($[P]=0.996; P_{min}=0.14$), ($[P]=0.997; P_{min}=0.07$), and
($[P]=0.997; P_{min}=0.18$) respectively. It appears that at this
level of precision the increase of the numbers of local moves is
of minor importance for the production of true GSs of the
spin-glass system, or to put it differently, it appears that the
production of true GSs depend only on the product $nt_{av}$. At
this point, one should also appreciate that the advertised, in
previous sections, increase of efficiency by the increase of the
numbers of local moves is compensated by the fact that one can
increase the above probabilities by increasing also the number of
independent runs ($N_r$) for each sample, as shown by
Roma~\emph{et al.}~\cite{roma09}.

Finally, we report here that using the $[CAE,BKL,(nn)_a]$, of the
second group in Table~\ref{tab:3}, we have estimated the sample
average of the GS energy per site, for a larger set of $N_s=10000$
samples, to be
$u_6=\sum\limits_{j=1}^{N_s}u_{6,j}/N_s=-1.7711(6)$. This is now
comparable with the estimate $u_6=-1.7714(3)$ given by
Roma~\emph{et al.}~\cite{roma09} for a sample set with the same
number of disorder realizations ($N_s=10^4$).

\subsection{Ground state energy of the anisotropic 3D EAB model}
\label{sec:4b}

\begin{table*}
\caption{\label{tab:4} PT parameters and GS energy per site of the
3D anisotropic EAB model. The last column is the difference of GS
energies per site between the isotropic, from Table B.3 of
Ref.~\cite{roma09} and the present anisotropic case. All runs were
carried out by the $[CAE,BKL,(nn)_a]$ protocol using an initial
part $t_{eq}=2N$.}
\begin{ruledtabular}
\begin{tabular}{lccccccc}
    $L$  &$(r;M)$ &$n=N_{local}/(M-1)$ &$N_s$ &$t_{av}$ &$u_L(ani)$ &$u_L-u_L(ani)$ \\
\hline

                               $3$  & (0.5;5) &$2$ &$5\times10^5$ &$2.7\times10^2$ &-1.7642(3) &0.0925\\
                               $4$  & (0.35;5) &$4$ &$10^5$ &$6.4\times10^2$ &-1.7703(3) &0.0328\\
                               $5$  & (0.2;5) &$9$ &$10^5$ &$1.25\times10^3$ &-1.7733(3) &0.0122\\
                               $6$  & (0.1;5) &$216$ &$2\times10^4$ &$3.24\times10^3$ &-1.7762(3) &0.0048\\
                               $7$  & (0.15;7) &$17$ &$6\times10^3$ &$3.43\times10^4$ &-1.7786(3) &0.0014\\
                               $8$  & (0.1;7) &$102$ &$10^4$ &$4.2\times10^4$ &-1.7803(3) &0.0003\\
                               $9$  & (0.1;9) &$27$ &$2\times10^3$ &$1.2\times10^6$ &-1.7825(3) &0.0001\\
                               $10$  & (0.1;11) &$120$ &$2.5\times10^3$ &$1.2\times10^6$ &-1.7830(2) &0.0000\\
                               $12$  & (0.1;13) &$43$ &$10^2$ &$10^7$ &-1.7850(8) &0.0001\\
                               $14$  & (0.1;13) &$68$ &$10^2$ &$1.4\times10^7$ &-1.7862(8) &0.0004\\
\end{tabular}
\end{ruledtabular}
\end{table*}

\begin{figure}[htbp]
\includegraphics*[width=9 cm]{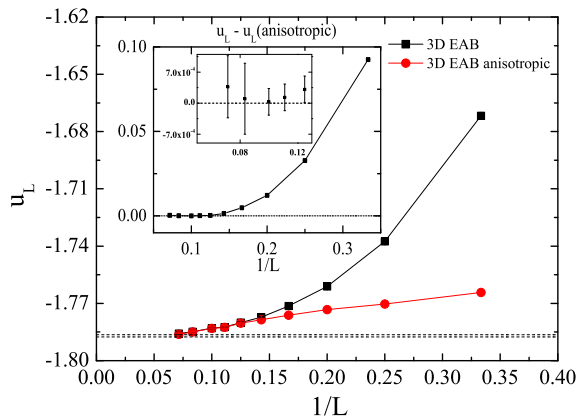}
\caption{\label{fig:10}(Color online) Finite-size behavior of GS
energies per site for the 3D EAB model and the present anisotropic
model. In the insets we show their difference, which as shown in
the nested inset, is much smaller than the estimated errors for
$L\geq6$.}
\end{figure}

We now consider the finite-size behavior of the GS energy of the
anisotropic case $(p_z=0,p_{xy}=0.5)$. As mentioned earlier, the
anisotropic model, $p_z=0 ; p_{xy}\leq \frac{1}{2}$ with
$J^z=J^{xy}=J(=1)$, has been studied recently by the present
authors, and its phase diagram has been presented in some
detail~\cite{anis_2}. The irrelevance of the anisotropy for the
ferromagnetic - paramagnetic transitions was established, and
further signs of universality concerning the other two kinds of
transitions, ferromagnetic-spin glass and spin glass-paramagnetic,
were pointed out. Of particular interest, for our study here, is
the observation~\cite{anis_2} that, the phase diagram points of
the spin glass-paramagnetic transition for the isotropic
$(p_z=p_{xy}=0.5)$ and the anisotropic $(p_z=0,p_{xy}=0.5)$ cases
are very close or even coincide. However, such a coincidence is a
prediction which, although appealing, goes well beyond the general
universality question, and cannot be trusted before a formal proof
is provided. We will now present results, that indicate an
analogous situation for the asymptotic limit of the GS energy of
these two cases, increasing the interest on a possible equivalence
in the asymptotic limit.

Following the approach of~\cite{roma09}, we produce here estimates
for the finite-size GS energy per site for the anisotropic model.
Our simulations cover the range of sizes $L=3-14$ and we are using
the PT protocol $[CAE,BKL,(nn)_a]$, which is a simple and
efficient choice. In all runs, we choose $N_r=1$, use a short
disregarding part ($t_{eq}=2N$) and vary the rest of the PT
parameters as shown in Table~\ref{tab:4}. In particular, the main
running times ($t_{av}$) are almost comparable to those in Table
B.3 of~\cite{roma09}. The temperature range used is approximately
the range $T=0.4-2.0$ and the CAE T-sequences were obtained using
the practice outlined in the previous section, with an exchange
rate, and corresponding number of temperatures, indicated in
second column of Table~\ref{tab:4}. For some sizes ($L=6,8$, and
$L=10$), the numbers of local moves were varied in alternative
runs, to test their effect on the estimated GS energies. Our
estimates of GS energies for the anisotropic model are given in
Table~\ref{tab:4}. Also, in the last column of the table, we give
the differences of GS energies per site between the isotropic
model (from Table B.3 of Ref.~\cite{roma09}) and the present
anisotropic model.

In Fig.~\ref{fig:10}, we show the finite-size behavior of the GS
energy per site for both isotropic and anisotropic 3D EAB models.
In the insets, their differences are illustrated. In particular,
it shown that their differences, for $L\geq6$ are much smaller
than the estimated errors. Therefore, the asymptotic limit of
these GS energies practically coincide. The two dashed lines in
the main panel indicate previous asymptotic estimations, namely
$u_{\infty}=-1.7863(4)$~\cite{pal96} and
$-1.7876(3)$~\cite{hart97}.

\section{Summary and Conclusions}
\label{sec:5}

We reviewed several PT schemes and examined their accuracy and
efficiency. Our tests on the 2D Ising model suggest that, the two
different methods of selecting the temperature sequences (CAE and
CEI) considered in this paper, produce results that are accurate
and are almost equivalent in efficiency. Efficiency of PT
protocols is greatly increased by using numbers of local moves
related to the canonical correlation times, as proposed earlier by
Bittner~\emph{et al.}~\cite{bittner08}. Accordingly, we found that
PT protocols using a Wolff algorithm for the local moves increase
the efficiency of the schemes. In particular an all-pair exchange
method, the $APE_B$ of Brenner \emph{et al.}~\cite{brenner07},
when used with local Wolff updates, has been found reasonably
accurate and very efficient. However, it was also found that, in
general, $APE_B$ protocols may show an unreliable behavior in
estimating thermodynamic equilibrium properties, such as the
specific heat behavior illustrated in Fig.~\ref{fig:2}. As argued,
this may be related to an improper implementation of the kinetic
MC method, that avoids the use of time weights for the MC
averaging process.

We also considered the problem of GS production of the 3D EAB
model, and we demonstrated the performance and relative efficiency
of several PT protocols. We found that PT protocols based on the
CAE T-sequences appear to be slightly better than those based on
the corresponding CEI T-sequences. In all our tests, the
superiority of the PT protocols involving BKL (or n-fold way)
local updates was firmly established. Finally, we presented
evidence that the asymptotic limit of the GS energy of the
isotropic $(p_z=p_{xy}=0.5)$ and the anisotropic
$(p_z=0,p_{xy}=0.5)$ EAB models, are very close, and possibly
coincide. This seems relevant to an analogous interesting behavior
found recently for the finite temperature phase diagram points
between spin-glass and paramagnetic phases~\cite{anis_2}.

\begin{acknowledgments}
T.P has been supported by a Ph.D grant of the Special Account of
the University of Athens.
\end{acknowledgments}

{}

\end{document}